\documentclass[4paper,twocolumn]{article}
\usepackage{amsmath} 
\usepackage{pstricks}
\usepackage{longtable}
\usepackage{graphicx}
\usepackage{epstopdf} 
\usepackage{color}
\usepackage{cite}
\usepackage{amssymb}
\input{epsf} 
\newcommand\f[2]{\frac{#1}{#2}} 
\def\beq{\begin{equation}} 
\def\eeq{\end{equation}} 
\def\to{\rightarrow} 
\def\nn{\nonumber} 
\def\beeq{\begin{eqnarray}}
\def\eeeq{\end{eqnarray}}

\newcommand\as{\alpha_{\mathrm{S}}}

\def\b0{b_0}

\def\gg{\gamma \gamma}
\def\Mgg{M_{\gamma \gamma}}
\def\to{\rightarrow}
 
\def\nn{\nonumber}

\def\mures{\mu_{res}}

\newcommand{\bea}{\begin{eqnarray}}
\newcommand{\eea}{\end{eqnarray}}

\newcommand{\ba}{\begin{eqnarray}}
\newcommand{\ea}{\end{eqnarray}}

\def\Ag4{{A_{g,4}}}
\def\Dg4{{D_{g,4}}}

\def\pqq(#1){p_{qq}(#1)}
\def\pgg(#1){p_{gg}(#1)}
\def\H(#1){{\rm{H}}_{#1}}
\def\Hh(#1,#2){{\rm{H}}_{#1,#2}}
\def\THh(#1,#2){{\widetilde{\rm{H}}}_{#1,#2}}
\def\Hhh(#1,#2,#3){{\rm{H}}_{#1,#2,#3}}
\def\THhh(#1,#2,#3){{\widetilde{\rm{H}}}_{#1,#2,#3}}

\def\beq{\begin{equation}}
\def\eeq{\end{equation}}
\def\beeq{\begin{eqnarray}}
\def\eeeq{\end{eqnarray}}
\def\nn{\nonumber}

\newcommand{\customlabel}[2]{%
\protected@write \@auxout {}{\string \newlabel {#1}{{#2}{}}}}
\def\eqalign#1{\null\,\vcenter{\openup\jot\m@th 
  \ialign{\strut\hfil$\displaystyle{##}$&$\displaystyle{{}##}$\hfil 
      \crcr#1\crcr}}\,}

\def\spa#1.#2{\left\langle#1\,#2\right\rangle}
\def\spb#1.#2{\left[#1\,#2\right]}

\begin{document} 

\begin{titlepage}
\renewcommand{\thefootnote}{\fnsymbol{footnote}}
\begin{flushright}
ICAS 29/17 
\end{flushright}
\par \vspace{10mm}

\begin{center}
{ \Large \bf 
Transverse-momentum resummation for the signal-background interference in the $H \to \gamma\gamma$ channel at the LHC}
\end{center}
\par \vspace{2mm}
\begin{center}
{\bf L. Cieri $^ {a}$}\footnote{lcieri@physik.uzh.ch},
{\bf F. Coradeschi $^ {b}$}\footnote{fc435@cam.ac.uk},
{\bf D. de Florian$^ c$}\footnote{deflo@unsam.edu.ar}\footnote{On leave of absence from Departamento de F\'\i sica, FCEyN, Universidad de Buenos Aires},
{\bf N. Fidanza$^ c$}\footnote{nfidanza@df.uba.ar}

\vspace{5mm}

$^a$ Institut f\"ur Theoretische Physik, Universit\"at Z\"urich, 
CH-8057 Z\"urich, Switzerland

$^b$ Department of Applied Mathematics and Theoretical Physics (DAMTP)\\ University of Cambridge, Centre for Mathematical Sciences, Wilberforce Road, Cambridge CB3 0WA

$^c$ International Center for Advanced Studies (ICAS), ECyT-UNSAM, \\ Campus 
Miguelete, 25 de Mayo y Francia, (1650) Buenos Aires, Argentina

\vspace{5mm}

\end{center}

\par \vspace{2mm}
\begin{center} {\large \bf Abstract}\end{center}

We present an upgraded calculation of the effects of resonance-continuum interference for the Higgs boson decaying to two photons at the Large Hadron Collider, at next-to-leading order in the strong coupling $\as$, $\mathcal{O}(\as^3)$, and including transverse-momentum ($q_T$) resummation at next-to-leading logarithmic accuracy. We study the importance of the interference contribution in different transverse-momentum regions, with a  particular focus on the low $q_T$ region $q_T^2 << Q^2$ (with $Q^2$ being the invariant diphoton mass) where resummation becomes essential for a reliable calculation.

\vspace*{\fill}
\begin{flushleft}
June 2017
\end{flushleft}
\end{titlepage}

\setcounter{footnote}{1}
\renewcommand{\thefootnote}{\fnsymbol{footnote}}

\section{Introduction}
The discovery of a new boson with a mass of approximately 125 GeV~\cite{Chatrchyan:2012xdj,Aad:2012tfa} at the Large Hadron Collider (LHC), having properties consistent with those predicted for the Standard Model (SM) Higgs boson, has opened a new era in precision particle physics. For the first time we have the possibility of directly probing the nature of the electroweak symmetry breaking mechanism and, possibly, completing our experimental verification of the SM. Considering that any new physics related to electroweak breaking (in particular, naturalness-inspired extensions of the SM such as supersymmetry or Higgs compositeness) will typically shift Higgs-related observables from the predicted SM values, one of the highest priorities nowadays is to improve both our theoretical predictions for, and experimental measures of, Higgs properties as accurately as possible, trying to uncover, or constrain as much as possible, any deviations from the SM.

In the context of LHC, the production of diphoton pairs offers a very clean experimental signature. Not only was this crucial for the Higgs discovery, it also provides an ideal scenario for probing its properties. In the inclusive case, i.e. the resonant process $pp \rightarrow H (\rightarrow \gamma \gamma) + X$, in the SM, the main production channel is gluon fusion, while the associated background has large contributions in both $q \bar{q}$ and $gg$ channels, with the latter formally subdominant but still sizable numerically. Since the Higgs is a narrow resonance, the interference between the Higgs boson signal and the continuum background (first appearing in the $gg$ channel) is suppressed and thus is often neglected in calculations. However, with measurements consistently improving in precision, it is increasingly important to have SM theoretical predictions which are as accurate as possible, to make sure that all relevant effects are included. In this paper we will focus on giving a state-of-the-art evaluation of interference effects in the diphoton channel in the SM.

In studying interference effects, it is useful in general to divide its contribution to observable distributions into two components, proportional to the real and imaginary parts of the Higgs boson Breit-Wigner propagator respectively (which by a slight abuse of language we will just call ``real'' and ``imaginary'' parts in the following). The diphoton invariant-mass distribution for the real part is odd around the Higgs mass, and as such contributes negligibly to the experimentally observed cross-section, which is integrated over the size of a bin ($\sim 1$ GeV), always much larger than the narrow theoretical lineshape ($\Gamma_H \simeq 4.2$ MeV in the SM). The imaginary part by contrast is even around the Higgs mass and can interfere constructively or destructively with the signal contribution. As has been known for a long time~\cite{Dicus:1987fk}, for a light SM Higgs boson ($m_H < 2 m_t$) the imaginary part vanishes at the leading-order (LO) in the strong coupling constant $\as$ in the limit of vanishing quark masses (for all quarks except the top). At higher orders~\cite{Dixon:2003yb} the dominant contribution to the total cross-section comes from the two-loop $gg \rightarrow \gamma \gamma$ amplitude, which gives a destructive interference leading to an order percent suppression of the total rate. Notice that despite the smallness of the suppression, its dependence on the Higgs width $\Gamma_H$, with the increased precision measurements from the LHC run 2, could be useful to provide a constraint on deviations of $\Gamma_H$ from the SM value, as pointed out recently in ref.~\cite{Campbell:2017rke}.

Even though the real part of the interference does not contribute significantly to the total cross-section, its impact may be important for other observables. In particular, as first pointed out by Martin~\cite{Martin:2012xc}, already at LO the real part introduces a shift in the position of the Higgs boson peak in the diphoton invariant mass distribution (and thus in the observed Higgs mass in the diphoton channel), which could be potentially experimentally detectable at the LHC. This happens because of an interplay between theoretical and experimental effects: the purely theoretical invariant mass distribution already shows a tiny shift in the position of the peak, due to interference, which is of order of the Higgs width. When the invariant mass distribution is enlarged by convolution with the order GeV detector resolution, the shift of the peak is also magnified, in a way which is roughly proportional to the resolution itself. A complete analysis of this effect to $\mathcal{O}(\as^2)$ was presented in ref.~\cite{deFlorian:2013psa} and further studied in ref.~\cite{Martin:2013ula}: the magnitude of the effect in LHC experimental conditions is around $\sim 100$ MeV, which is of the same order~\cite{Aad:2015zhl} of the current experimental resolution, suggesting that indeed the effect could be detectable in the near future. The analysis was extended to NLO in ref.~\cite{Dixon:2013haa}, showing a reduction with respect to the LO prediction by about $30\%$ (due essentially to the different K-factors of the signal and of the interference contributions) which is significant but not enough to make the effect negligible.

It is worth noting that the absolute value of the predicted mass shift at NLO could have a rather strong dependence on the Higgs transverse-momentum $q_T$~\cite{Dixon:2013haa}, at least in presence of real radiation. This $q_T$ dependence may allow for the direct measurement of the mass shift just using diphotons without reference to other channels (e.g. ZZ, where the shift is expected to be much lower on theoretical grounds). Furthermore, the bulk of the events is produced at relatively low $q_T$ where large logarithmic terms appear in the cross-section spoiling the convergence of the perturbative series. Therefore to asses the robustness of the prediction, and to get a reliable estimate of its $q_T$ dependence, it is essential to consider transverse-momentum resummation for the Higgs boson $q_T$, as was also noted in refs.~\cite{Dixon:2013haa,Becot:2015xzw}. 

The aim of this paper is to upgrade the evaluation of the effect of interference (from both real and imaginary parts) in the diphoton channel at NLO ($\mathcal{O}(\as^3)$) by including resummation of transverse-momentum logarithmically-enhanced contributions, thus ensuring the reliability of our predictions also in the small $q_T$ region.

\section{Theory overview}

The Higgs boson production cross-section in the gluon fusion channel, at the completely differential level, is known to the next-to-next-to-leading-order (NNLO) in perturbative QCD~\cite{Catani:2007vq,Anastasiou:2004xq,Anastasiou:2005qj}. Also well known are the results including transverse-momentum ($q_T$) resummation to the next-to-next-to-leading-logarithmic (NNLL) accuracy~\cite{deFlorian:2011xf}. The diphoton background was calculated to the NNLO in QCD first in ref.~\cite{Catani:2011qz} and later in ref.~\cite{Campbell:2016yrh}. The corresponding $q_T$-resummed predictions, to NNLO+NNLL accuracy, were published in ref.~\cite{Cieri:2015rqa}.

Our strategy here will be to combine the known fixed-order results up to the dominant NLO ($\mathcal{O}(\as^3)$) for signal-background interference~\cite{Dixon:2013haa} with transverse-momentum resummation to NLL accuracy (the $\mathcal{O}(\as^4)$ prediction for the background in the $gg$ channel, which would be needed to go to NNLO in the interference, is still not known). The Higgs propagator is approximated via a simple Breit-Wigner shape, using $m_H = 125$ GeV and $\Gamma_H = 4.2$ MeV as the theoretical Higgs mass and width respectively~\cite{Djouadi:1997yw}, and diagrams where the Higgs appears inside a loop are neglected.
In any given partonic channel $a_1 a_2$, the scattering amplitude to two photons can be written as
\begin{equation}
\mathcal{A}_{a_1 a_2 \to \gg} = \frac{\mathcal{A}_{sig}}{(M_{\gg}^2 - m_H^2) + i \Gamma_H m_H} + \mathcal{A}_{bkg} \, ,
\end{equation}
where $M_{\gg}$ is the diphoton invariant mass. The signal and background subamplitudes $\mathcal{A}_{sig}$ and $\mathcal{A}_{bkg}$ are functions of  $M_{\gg}$. 
The residue of the amplitude $\mathcal{A}_{a_1 a_2 \to \gg}$ at the complex Higgs pole $M_{\gg}^2 = m_H^2 + i m_H \Gamma_H$ is a well-defined and gauge-invariant quantity, which we can identify with $\mathcal{A}_{sig}(m_H^2 + i m_H \Gamma_H) \simeq \mathcal{A}_{sig}(m_H^2)$. This analytic structure is shared by all channels, so that when we square the amplitudes and combine them to build the diphoton invariant mass distribution $\frac{d \sigma^{\gg}}{d M_{\gg}}$ we obtain
\begin{equation}
\label{BSRI}
\begin{split}
\frac{d \sigma^{\gg}}{d M_{\gg}} & = B + \frac{S}{(M_{\gg}^2 - m_H^2)^2 + \Gamma_H^2 m_H^2}\\
& + \frac{(M_{\gg}^2 - m_H^2) R + \Gamma_H m_H I}{(M_{\gg}^2 - m_H^2)^2 + \Gamma_H^2 m_H^2} \; ,
\end{split}
\end{equation}
where $B$, $S$, $R$ and $I$ (with, in particular, $B \propto \sum |\mathcal{A}_{bkg}|^2$, $S \propto \sum |\mathcal{A}_{sig}|^2$, $R \propto \sum \text{Re}(\mathcal{A}_{sig}\mathcal{A}_{bkg}^*)$ and $I \propto \sum \text{Im}(\mathcal{A}_{sig}\mathcal{A}_{bkg}^*)$, the sum running over the various partonic channels) are -- around the Higgs pole -- slowly varying functions of $M_{\gg}$, which we will simply call background, signal and real and imaginary interference contributions to the differential cross-section in the following.

When integrating over the Higgs peak, the most important contribution comes from the signal ($\pi S/(2m_H^2 \Gamma_H)$), with a small correction from the imaginary part of the interference ($\pi I/(2m_H)$). The contribution of the background $B$ can be \textit{identificated} (predicted or fitted) and thus neglected, while the contribution of the real part of the interference $R$ is averaged to $0$ due to its (approximately) antisymmetric dependence on $M_{\gg}^2- m_H^2$. The main effect of $R$ is instead~\cite{Martin:2012xc} to induce a slight shift of position of the peak of the distribution, and consequently of the \emph{observed} Higgs mass $m_H^{obs}$. The precise magnitude of the \emph{mass shift} $\Delta m_H \equiv m_M^{obs} - m_H$ depends mainly on the detector resolution $\sigma_{exp}$ but also, evidently, on the details of the experimental fit procedure, whose exact modeling is beyond the scope of the present work. Here we will give an estimate of the shift using the procedure described in ref.~\cite{Dixon:2013haa} for $\Delta m_H$. In general we will use results of~\cite{Dixon:2013haa} as a direct comparison\footnote{Notice that the kinematical cuts are not completely equivalent in both analyses, as it is stated in Section~\ref{sec:NumRes}.} for all our results involving $\Delta m_H$. For this reason, we also use the same representative value for the experimental width as in~\cite{Dixon:2013haa}, namely $\sigma_{exp} = 1.7$ GeV.

\begin{figure}[!hbt]
\begin {center}
\includegraphics[width=0.5\textwidth]{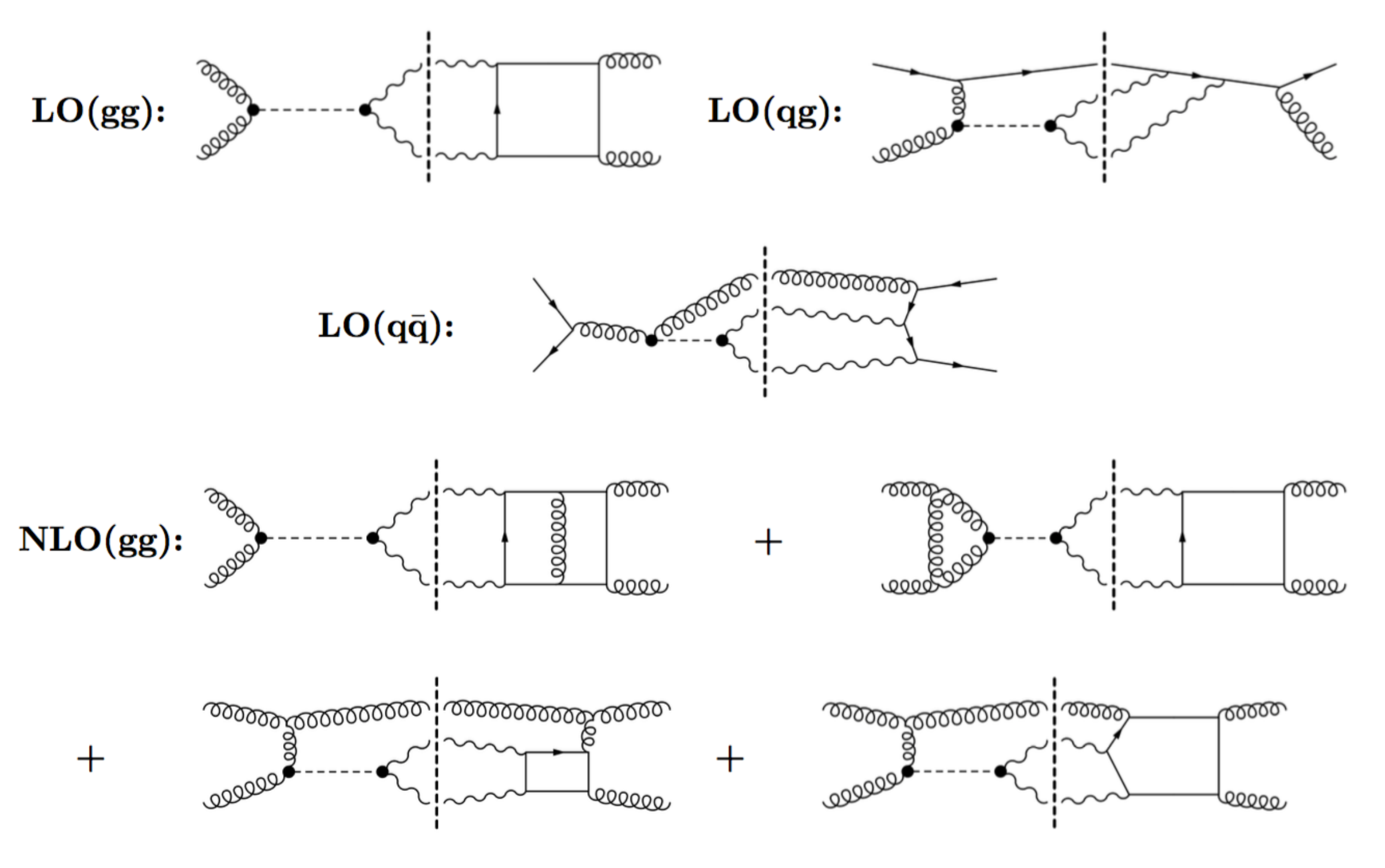}
\hspace{0.5cm} 
\caption{\label{fig:1} \small{Representative Feynman diagrams for the different contributions to interference between the Higgs resonance and the continuum background in the diphoton channel up to $\mathcal{O}(\as^3)$.}}
\end {center}
\end{figure}
In figure~\ref{fig:1} we schematically show the different partonic channels and perturbative orders which enter in our calculation. The dominant LO ($\mathcal{O}(\as^2)$) contribution is given by the interference of the resonant amplitude $gg \rightarrow H \rightarrow \gamma \gamma$ with the one-loop continuum $gg \rightarrow \gamma \gamma $ amplitude mediated by the five light quark flavors; this is the contribution originally included in the analysis of ref.~\cite{Dicus:1987fk}. Two other channels technically contribute at LO: the tree-level processes $qg \rightarrow \gamma \gamma q$ and $q\bar{q} \rightarrow H g \rightarrow \gamma \gamma g$, where real QCD radiation is produced alongside the two photons. These additional contributions were first considered in refs.~\cite{deFlorian:2013psa,Martin:2013ula} and although formally of the same order as the $gg$ channel one, they are actually suppressed by the smaller quark/antiquark PDFs. We include them anyway for the sake of completeness.

At NLO ($\mathcal{O}(\as^3)$) we only include the dominant contribution, the $gg$ channel one. It is made up of three different kind of amplitudes that are schematically depicted in the third and fourth rows of fig.~\ref{fig:1}: the virtual corrections to $gg \rightarrow H$ production and to the $gg \rightarrow \gamma \gamma$ box contribution to the background (a two loop effect) respectively, plus the real radiation correction to the LO $gg$ channel, in which the tree-level amplitude $gg \rightarrow H g \rightarrow \gamma \gamma g$ interferes with the real correction to the box contribution to the continuum background. All these amplitudes are taken and adapted from refs.~\cite{Bern:1993mq,Bern:1994fz,Spira:1995rr,Bern:2001df}.

As sketchily depicted in figure~\ref{fig:1}, we consider the effective Higgs coupling to gluons in the limit of a very heavy top quark. This $m_t \rightarrow \infty$ effective theory for the 
Higgs interactions with gluons constitutes a good approximation~\cite{Shifman:1979eb,Dawson:1990zj,Djouadi:1991tka,Dawson:1993qf} for the exact case ($m_t = 172.5$ GeV), for transverse momenta of the Higgs boson satisfying  $q_{T}^H < m_t$.
The Higgs interaction with the two photons is instead treated using the complete one-loop expression:
\begin{eqnarray}
C_{\gamma} &= &-\frac{\alpha M_{\gg}^2}{4\pi v} \biggl [
F_1(4 m_W^2/M_{\gg}^2) \\\nonumber
&+& 
\sum_{f = t,b,c,\tau} N_{c}^f e_f^2 F_{1/2} (4 m_f^2/M_{\gg}^2) \biggr ],
\label{eq:CgammaH}
\end{eqnarray}
where $v\approx 246$ GeV is the Higgs expectation value, $\alpha = 1/137$ , $N_c^f=3$ (1) for $f=$ quarks (leptons) with electric charge $e_f$ and mass $m_f$, and
\begin{eqnarray}
F_1(x) &=& 2 + 3 x [1 + (2-x) f(x)],\nonumber
\\
F_{1/2} (x) &=& -2 x [1 + (1-x) f(x)],\nonumber
\\
f(x) &=& \Biggl \{ \begin{array}{ll}
[\arcsin (\sqrt{1/x})]^2, & x\geq 1 ,
\\\nonumber
-\frac{1}{4} \left [ \ln \left (\frac{1 + \sqrt{1-x}}{1 - \sqrt{1-x}}\right ) 
- i \pi \right ]^2,\,\,
& x\leq 1.\nonumber
\end{array}
\end{eqnarray}
For the continuum amplitude we use the five flavor massless box (or pentagon) approximation~\cite{Bern:2002jx}.

We combine the various contributions using the transverse-momentum resummation formalism developed in~\cite{Bozzi:2005wk,Catani:2013tia} (and references therein), which allows us to take into account logarithmically-enhanced contributions to all orders in $\as$ while also implementing the cancellation of infrared singularities between real and virtual corrections to the Born level processes. We highlight briefly the features of the formalism most relevant to the present calculation, referring to the original papers for further details. 

The starting point is a decomposition of the partonic cross-section (fully differential in the diphoton variables) into a ``resummed'' and a ``finite'' term
\begin{equation}
\label{resplusfin}
\f{d{\hat \sigma}^{\gg}_{ab}}{dq_T^2}=
\f{d{\hat \sigma}_{\gg\,ab}^{(\rm res.)}}{dq_T^2}
+\f{d{\hat \sigma}_{\gg\,ab}^{(\rm fin.)}}{dq_T^2}\; ,
\end{equation}
where $a,b = q,\bar{q},g$ specify the different partonic channels. The distinction between the two terms on the right-hand side of eq.~\eqref{resplusfin} is purely theoretical. As already noted, the cross-section is differential in all photonic variables, namely the diphoton transverse-momentum $q_T$, invariant mass $M_{\gg}$, and pseudorapidity $y$ as well as two more variables needed to define the individual photons momenta, collectively denoted as $\Omega$; we will write the differential cross-section simply as $d\sigma_{\gg}/dq_T^2$ in the formulas, and as $d\sigma_{\gg}$ in the text, for the sake of brevity). The decomposition of $d{\hat \sigma}^{\gg}_{ab}$ in eq.~\eqref{resplusfin} is arbitrary to some extent: the main point is that $d{\hat \sigma}_{\gg\,ab}^{(\rm res.)}$ must collect all contributions which are singular as $q_T \to 0$ (in particular, the logarithmically-enhanced ones plus the Born and all purely virtual corrections, which are proportional to $\delta(q_T^2)$). Terms which are constant as $q_T \to 0$ can be included either in $d{\hat \sigma}_{\gg\,ab}^{(\rm res.)}$ or $d{\hat \sigma}_{\gg\,ab}^{(\rm fin.)}$; this choice is conventional and does not affect the full cross-section $d{\hat \sigma}^{\gg}_{ab}$ (in particular, the formalism is carefully built in such a way as to avoid any double-counting).

In more detail, the resummed component $d{\hat \sigma}_{\gg\,ab}^{(\rm res.)}$ is defined as the Fourier transform of a form-factor ${\cal W}$ defined in impact parameter (denoted by $b$) space\footnote{Eq.~\eqref{resum} is a slightly simplified expression omitting a sum over various flavour contributions, all of which share the same structure. For a detailed discussion, we refer to~\cite{Bozzi:2005wk,Bozzi:2007pn}.}:
\begin{align}
\label{resum}
& \f{d{\hat \sigma}_{\gg \,ab}^{(\rm res.)}}{dq_T^2}(q_T,\Mgg,y,{\hat s},\Omega;
\as,\mu_R^2,\mu_F^2,\mures^2)
= \f{\Mgg^2}{\hat s} \\
& \cdot \int_0^\infty \!\!\!db \; \f{b}{2} \;J_0(b q_T) 
\;{\cal W}_{ab}^{\gg}(b,\Mgg,y,{\hat s},\Omega;\as,\mu_R^2,\mu_F^2,\mures^2) \;,\nn
\end{align}
where $J_0(x)$ is the $0$th-order Bessel function, and (only writing explicitly the most relevant variable dependences from now on)
\begin{equation}
\label{wtilde}
{\cal W}^{\gg}
 = {\cal H}^{\gg}(\as,\mures^2) \exp\{{\cal G}(\as,L,\mures^2)\} \; .
\end{equation}
The $b$ (and thus $q_T$) dependence in eq.~\eqref{wtilde} is confined to the logarithmic term $L\equiv \ln ({\mures^2 b^2}/{b_0^2})$, ($b_0\equiv2e^{-\gamma_E}$ where $\gamma_E$ is the Euler constant), only appearing in the coefficient function ${\cal G}$ which is defined by a perturbative expansion
\begin{equation}
\begin{split}
& {\cal G}(\as,L,\mures^2)= L \;g^{(1)}(\as L)\\
\label{exponent}
+ & \; g^{(2)}(\as L,\mures^2) + \; \as \, g^{(3)}(\as L,\mures^2) + \; \dots
\end{split}
\end{equation}
where the term $L\, g^{(1)}$ collects the leading logarithmic (LL) $\mathcal{O}(\alpha_s^{p+n}L^{n+1})$
contributions, the function $g^{(2)}$ includes the next-to-leading leading logarithmic (NLL) $\mathcal{O}(\alpha_s^{p+n}L^{n})$ contributions \cite{Kodaira:1981nh,Catani:1988vd}, $g^{(3)}$ controls the NNLL $\mathcal{O}(\alpha_s^{p+n}L^{n-1})$ terms \cite{Davies:1984hs, Davies:1984sp, deFlorian:2000pr,Catani:1989ne,Becher:2010tm} and so forth; $p$ is the number of powers of $\alpha_s$ in the LO (Born) process\footnote{The parameter $p$ in the case of the continuum diphoton background is $p_{\gg}=0$. In the case of the signal (Higgs boson production to two photons) $p_H=2$, and finally, in the case of the interference $p_{{\rm int.}}=2$.}. Furthermore the ${\cal G}$ factor is \emph{universal} (independent of the final state). The process dependence only appears in the ``hard'' coefficient ${\cal H^{\gg}}$, which is by contrast $b$-independent, does not contain any log-enhanced term and admits a standard perturbative expansion in $\as$. More specifically, ${\cal H^{\gg}}$ collects (along with some universal terms, see~\cite{Bozzi:2005wk} for its detailed structure) contributions from all the partonic subprocesses which can create the final state without any accompanying QCD radiation: in our specific case this is just $gg\to\gg$ for the interference ($q\bar{q}\to \gg$ can only contribute to the background, not to the signal). For each relevant subprocess, ${\cal H^{\gg}}$ includes the full Born term as well as contributions from all \emph{purely virtual} corrections, order by order in $\as$. These virtual contributions are extracted from the full corresponding matrix elements -- UV-renormalized but still IR-divergent, and regulated by dimensional regularization -- by applying a universal \emph{subtraction operator}, using the procedure described in detail in ref.~\cite{Catani:2013tia}. In our specific case, contributions to the NLO hard coefficient $\mathcal{H}^{(1)\,\gg}$ from Higgs signal, $\gg$ background and signal-background interference are all needed for the complete calculation; signal and background contributions are well-known and have been adapted from ref.~\cite{Catani:2013tia}, while the interference contribution was first explicitly calculated in work.

In eqs.~\eqref{resplusfin}-\eqref{wtilde} an auxiliary scale $\mures$ ($\mures\sim \Mgg$), the \emph{resummation scale} \cite{Bozzi:2005wk}, is also introduced. Its role is analogous to that of $\mu_R$ and $\mu_F$: although ${\cal W}^{\gg}$ does not (in principle) depend on $\mures$ when evaluated to all perturbative orders, a dependence appears when it is truncated at some level of logarithmic accuracy (see eq.~(\ref{exponent})), which can be used to estimate uncertainties introduced by the truncation.

Once the $d{\hat \sigma}_{\gg\,ab}^{(\rm res.)}$ is given, the second term in eq.~\eqref{resplusfin}, $d{\hat \sigma}_{\gg\,ab}^{(\rm fin.)}$, is then defined order by order in $\as$ by \emph{subtraction}:
\begin{equation}
\label{CT}
\left.\left(\f{d{\hat \sigma}_{\gg\,ab}^{(\rm fin.)}}{dq_T^2}\right)\right|_{f.o.} = 
\left.\left(\f{d{\hat \sigma}_{\gg\,ab}}{dq_T^2}\right)\right|_{f.o.} - 
\left.\left(\f{d{\hat \sigma}_{\gg\,ab}^{(\rm res.)}}{dq_T^2}\right)\right|_{f.o.} \!\!\!,
\end{equation}
where the term $d{\hat \sigma}_{\gg\,ab}$ collects (at NLO) the \emph{purely real}  corrections to the partonic cross-section, while the last term is the \emph{fixed-order} expansion of $d{\hat \sigma}_{\gg\,ab}^{(\rm res.)}$. The resummed term $d{\hat \sigma}_{\gg\,ab}^{(\rm res.)}$ is constructed in such a way that the left-hand side of eq.~\eqref{CT} is \emph{finite} as $q_T \to 0$, and with the requirement that the total cross-section be kept at the same value as in the standard fixed-order calculation. 


\section{Numerical results}
\label{sec:NumRes}

We now proceed to present our results for the resummed calculation, to NLO+NLL accuracy, of the interference between signal and background for a Higgs boson decaying to two photons at the LHC. We show in particular the relative importance of the contributions from the low $q_T$ (where the resummation is of utmost importance) and high $q_T$ regions.

The NLO calculation of the finite term $\left.\left(d{\hat \sigma}_{\gg\,ab}\right)\right|_{f.o.}$ (see eq.~\eqref{CT}) was implemented using the $2\gamma$NNLO \cite{Catani:2011qz} MonteCarlo code\footnote{The complete fixed-order calculation is implemented in $2\gamma$NNLO \cite{Catani:2011qz}.}. While the calculation of the resummed contribution $d{\hat \sigma}_{\gg\,ab}^{(\rm res.)}$ at NLL was implemented as in the 2$\gamma$Res~\cite{Cieri:2015rqa} code\footnote{The complete calculation including transverse-momentum resummation is included in 2$\gamma$Res~\cite{Cieri:2015rqa}.}.

Our numerical analysis is based on typical set of cuts used by recent Higgs boson searches and studies. In particular we impose $q_T^{\gamma~{\rm Hard}}\geq 40 \mbox{GeV}$, $q_T^{\gamma~{\rm Soft}}\geq 30 \mbox{GeV}$, and $|y_\gamma|\leq 2.5$. We rely on the smooth cone isolation criterion~\cite{Frixione:1998jh}:
\begin{align}\label{Eq:Isol_frixcriterion}     
&\sum E_{T}^{had} \leq E_{T \, max}~\chi(r)\;, \nn\\
&\mbox{\qquad inside any cone:} \\
& r^{2}=\left( y - y_{\gamma} \right)^{2} +    
\left(  \phi - \phi_{\gamma} \right)^{2}  \leq R^{2} \nn \;,    
\end{align}
with a standard choice for the function $\chi(r)$:
\begin{equation}
\label{Eq:Isol_chinormal}
\chi(r) = \left( \frac{1-\cos (r)}{1-\cos R} \right)^{n}\;,
\end{equation}
where we additionally set $n=1$ throughout the paper. The remaining isolation parameters are set to $E_{T~max}=3~$GeV and $R=0.4$. Finally, we require the minimum angular separation between the two photons to be $R_{\gg}=0.4$.

We set the center of mass energy at $\sqrt{s} = 8$ TeV, and we use the Martin-Stirling-Thorne-Watt (MSTW) 2008 NLO PDF set~\cite{Martin:2009iq}. Regarding scales, we set the central value of the renormalization and factorization scales to $\mu_R = \mu_F = m_H$, while for the resummation scale we choose $m_H/2$. The theoretical uncertainties due to the truncation of the perturbative series are estimated by performing standard variations of the renormalization, factorization and resummation scales. In particular, in order to obtain the uncertainty bands on all the plots shown in this Section, we consider asymmetric values for the renormalization and factorization scales and three choices for the resummation scale, for a total of six cases: $\mu_R = 2m_H$, $\mu_F = m_H/2$, $\mures = \mu_i$ and $\mu_R = m_H/2$, $\mu_F = 2 m_H$, $\mures = \mu_i$ with $\mu_i = m_H$, $m_H/2$ and $m_H/4$.

We performed numerous cross-checks on all the amplitude contributions used to build the present calculation. The real NLO contributions to the interference were checked by comparing, on the one hand, the signal with the Hq$_{{\rm T}}$~\cite{deFlorian:2011xf} code, and on the other hand the background with gamma2MC \cite{Bern:2002jx}.
We have numerically checked that the r.h.s. of eq.~\eqref{CT} is finite at the NLO. Finally, we checked that for large values of transverse-momentum ($q_T>2q_T^{\gamma~{\rm Hard}}$), where the resummation does not have a significant impact, the known fixed-order NLO results \cite{Dixon:2013haa} are recovered.

\begin{figure}[!hbt]
\begin {center}
\includegraphics[width=0.46\textwidth]{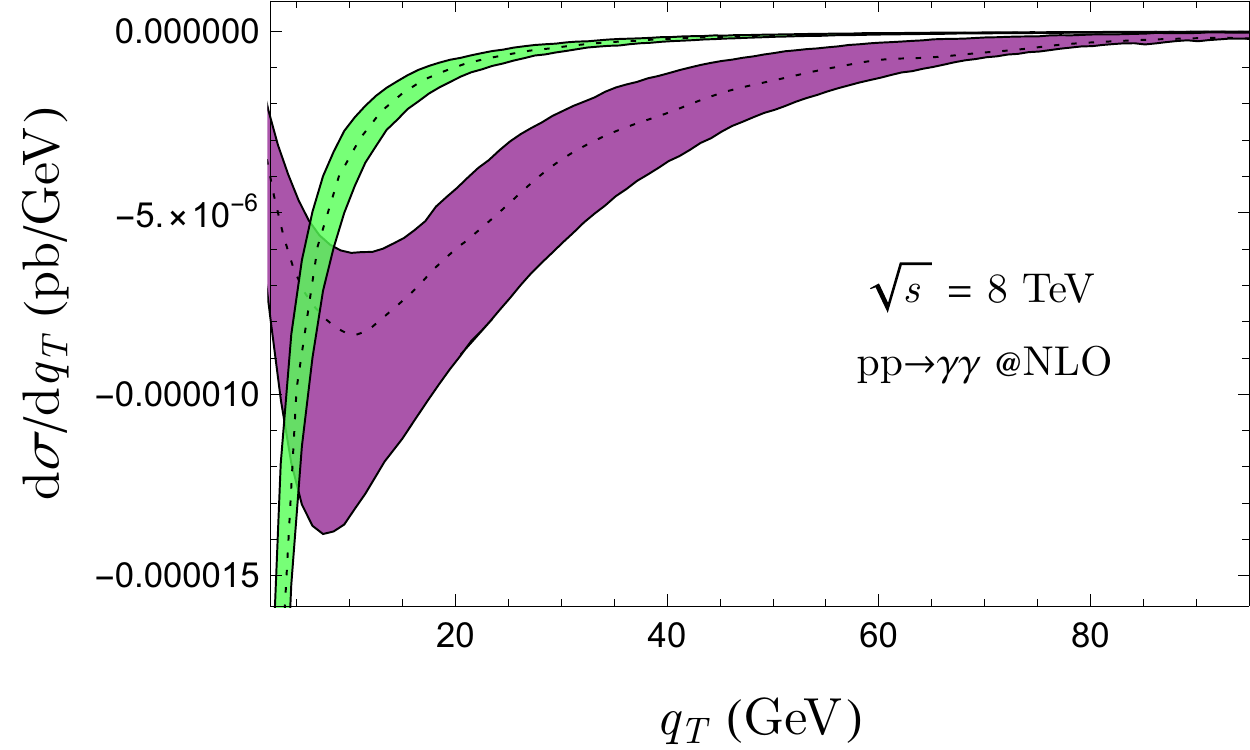}
\hspace{0.5cm} 
\caption{\label{fig:pTspectrumInterf} \small{Interference contribution to the differential cross-section in the transverse-momentum of the diphoton system at NLO (green curve) and NLO+NLL (violet curve). The dotted lines represent the result corresponding to the central scale choice $\mu_R=\mu_F=2\mures=m_H$, and the bands which show the theoretical uncertainties are obtained by considering asymmetric values for the renormalization and factorization scales and three choices for the resummation scale, for a total of six cases as described in the text.}}
\end {center}
\end{figure}

We now proceed to present our results, starting with the transverse-momentum distribution for the interference contribution which is given in figure~\ref{fig:pTspectrumInterf}, both at fixed-order (NLO, being this the lowest non trivial order for this distribution), as well as including transverse-momentum resummation at NLO+NLL.
We can see how the fixed-order distribution loses physical meaning in the low transverse-momentum region, where the enhanced logarithmic terms prevent the convergence of the perturbative expansion, causing the corresponding curve to diverge as $q_T \to 0$.
By contrast, the resummed calculation remains (as expected) well-defined down to very small values of $q_T$, which allows for reliable predictions for any physical quantity receiving important contributions from this region. This can be explained by noticing that transverse-momentum resummation \textit{redistributes} the weight of events that in the fixed-order calculation only enter at $q_T = 0$ over the entire $q_T$ space.
This consideration can also be extended to the uncertainty bands: at low $q_T$ for the fixed-order contribution, the uncertainty is grossly underestimated, while the resummed prediction provides a more reliable estimate of the errors. 

Two more effects  are noticeable at large values of $q_T$; the difference between the central values of the resummed and the fixed order calculations, and the much larger uncertainty band in the resummed calculation, caused by the large dependence on the resummation scale, that makes both calculations consistent. Both effects are originated mostly due to the size of the two-loop amplitude of the continuum ${\cal A}_{gg \rightarrow \gamma \gamma}^{(2)}$, present only in the resummed expression through the hard coefficient  $\mathcal{H}^{(1)\,\gg}$. More explicitly, the $q_T$ distribution involves an integration over invariant masses and, therefore, receives its main contribution from the imaginary part of the interference, which depends on the relative phase between the signal and continuum amplitudes. At the Born level the imaginary piece only arises from a rather small contribution of the production and decay components of the signal, and, eventually, by the heavy quark contribution in the one-loop continuum amplitude (not accounted here). On the other hand, a much larger imaginary part of ${\cal A}_{gg \rightarrow \gamma \gamma}$ occurs at the two-loop order, even for massless quarks in the loop. This contribution enhances the resummed expression considerably and, since at this order $\mathcal{H}^{(1)\,\gg}$ does not contribute to the expansion of the exponentiated contribution and the same level of accuracy has not been achieved in the fixed order result, it generates at the same time a large dependence on $\mures$ and a considerable enhancement in the distribution even at large transverse momentum. This situation anticipates that higher order contributions to the transverse momentum distribution, arising, for example, from two-loop amplitudes with extra real radiation, might also be very sizeable.

\begin{figure}[!hbt]
\begin {center}
\begin{tabular}{c}
\hspace*{-1cm} 
\includegraphics[width=0.46\textwidth]{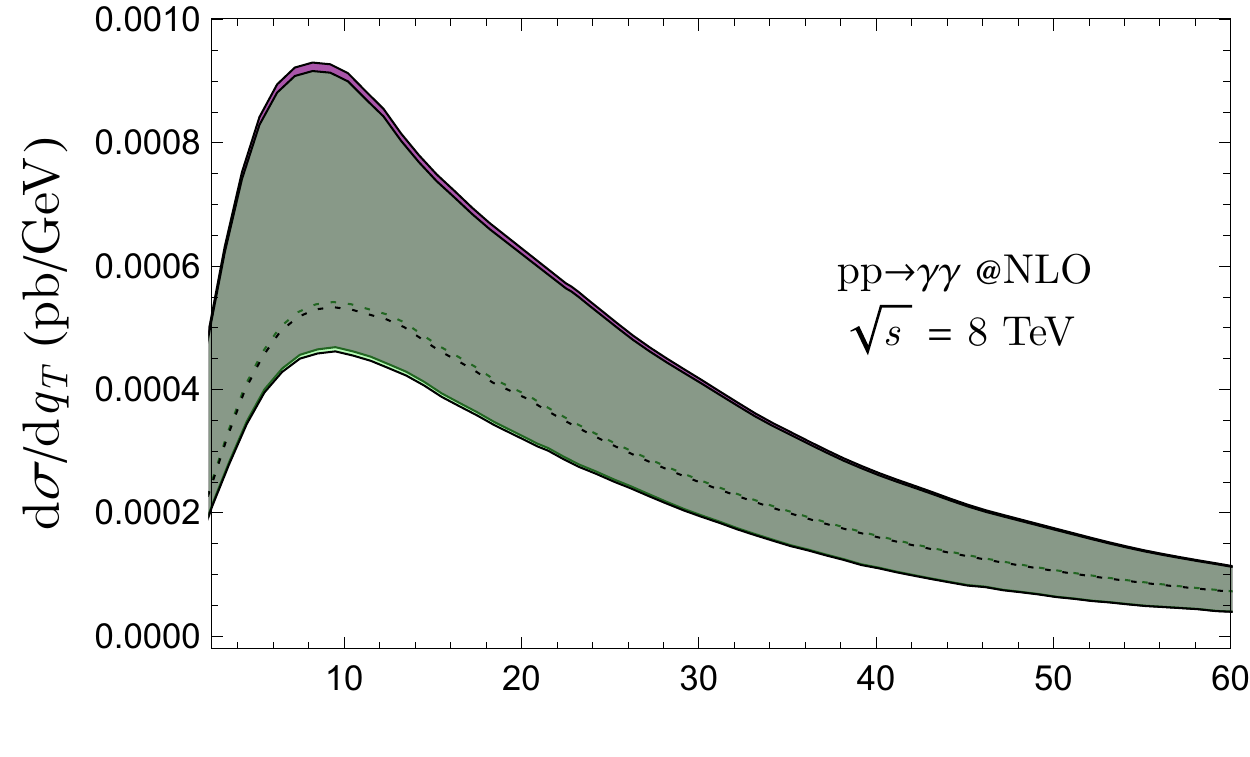} \vspace*{-0.5cm} \\
\hspace*{-1.02cm} 
\includegraphics[width=0.46\textwidth]{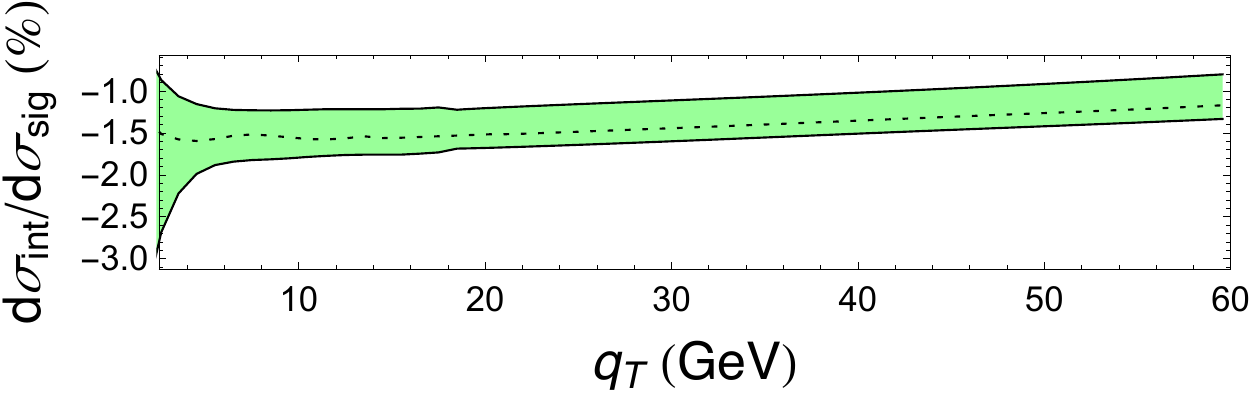}
\end{tabular}
\caption{\label{fig:pTspectrum} \small{Above: full transverse-momentum distribution of the diphoton pair at NLO+NLL (green curve) and the same observable without including the interference contribution (violet curve). The dotted lines represent the result corresponding to the central scale choice $\mu_R=\mu_F=2\mures=m_H$, and the bands show the theoretical uncertainties obtained as described in the text. As the interference contribution is small, the two curves are almost on top of each other. Below, we highlight the interference contribution by showing the ratio between the full result including signal plus interference, and the one just including the signal.}}
\end {center}
\end{figure}

In figure \ref{fig:pTspectrum} we show the transverse-momentum distribution for the signal alone (violet curve) versus the signal plus interference (green curve), both including resummation to NLO+NLL accuracy. The effect from including the interference terms is, as expected, quite small, resulting in a decrease of the differential cross-section (lower panel of figure \ref{fig:pTspectrum}) roughly of order $1\%$.

\begin{figure}[!hbt] 
\begin {center}
\hspace*{-0.65cm}
\includegraphics[width=0.48\textwidth]{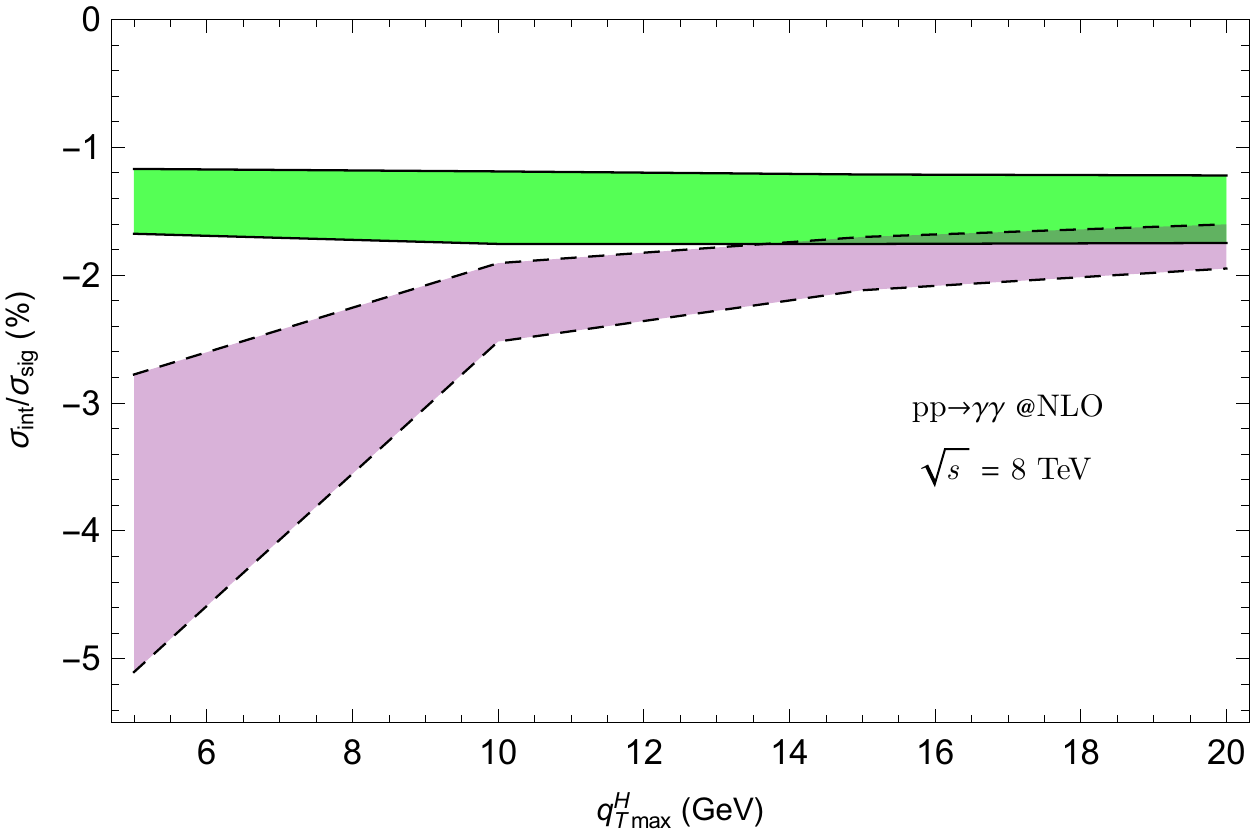}
\caption{\label{fig:destructive} \small{The percentual suppression of the total (peak) cross-section due to interference in terms of the maximum Higgs transverse-momentum $q_{T  max}^H$ for the calculation at NLO+NLL. The green band shows our resummed prediction at NLO+NLL and the band arising from variation of the scales as described at the beginning of this Section. The light purple band gives the corresponding prediction at fixed order, without resummation.}}
\end {center}
\end{figure}

It is interesting to observe precisely how the suppression of the cross-section by destructive interference depends on transverse-momentum: the $q_T$-dependence is in fact mild, ranging from about $1.5 \%$ at small $q_T$ to $\sim1\%$ at $q_T \gtrsim 50$ GeV, but anyway the effect is slightly larger at low $q_T$ values, that is in the region where resummation is more relevant. It is worth recalling that only the imaginary (symmetric around $m_H$) part of the interference $I$ (see eq.~\eqref{BSRI}) provides a significant contribution to the distributions in figures~\ref{fig:pTspectrumInterf} and~\ref{fig:pTspectrum}, since the real one $R$ (antisymmetric) is averaged out by the integration around $m_H$. The importance of resummation in evaluating the $q_T$ dependence of destructive interference is highligthed in figure~\ref{fig:destructive}.

To allow for a comparison with the fixed order calculation, we introduce a kinematic cut $q_{T  max}^H$ on the \emph{maximum} transverse-momentum $q_T$ of the diphoton pair (of the Higgs), that is, we only keep events which satisfy $q_T < q_{T  max}^H$ (the cut is imposed in addition to the common ones used for all our results and detailed at the beginning of this Section), thus emphasizing the contribution of the low-$q_T$ region, where resummation effects are most important. Note that, since the calculation is at NLO, a cut on $q_T$ is roughly equivalent to a cut on the transverse-momentum of an additional jet in the final state accompanying the diphoton system.

The resummed prediction corresponding to the $q_{T max}^H$ cut is shown in figure~\ref{fig:destructive} (green band). Again, we see that the suppression does not have a strong dependence on $q_T$, in accord to what was shown in figure~\ref{fig:pTspectrum}. This is significantly different from what is obtained by the corresponding fixed order calculation, where an apparent dependence on the $q_{T max}^H$ cut is evident. The difference is to be expected since NLO corrections at fixed-order are affected by large corrections at low $q_{T  max}^H$ due to the presence of large logarithmic terms in the cross-section. Our resummed calculation is by contrast stable, allowing for a reliable prediction also in the very small $q_{T  max}^H$ region. Notice that the two predictions start to converge at the rightmost edge of figure~\ref{fig:destructive}, that is at $q_{T max}^H \simeq 20$ GeV.

\begin{figure}[!hbt] 
\begin {center}
\hspace*{-0.65cm}
\includegraphics[width=0.48\textwidth]{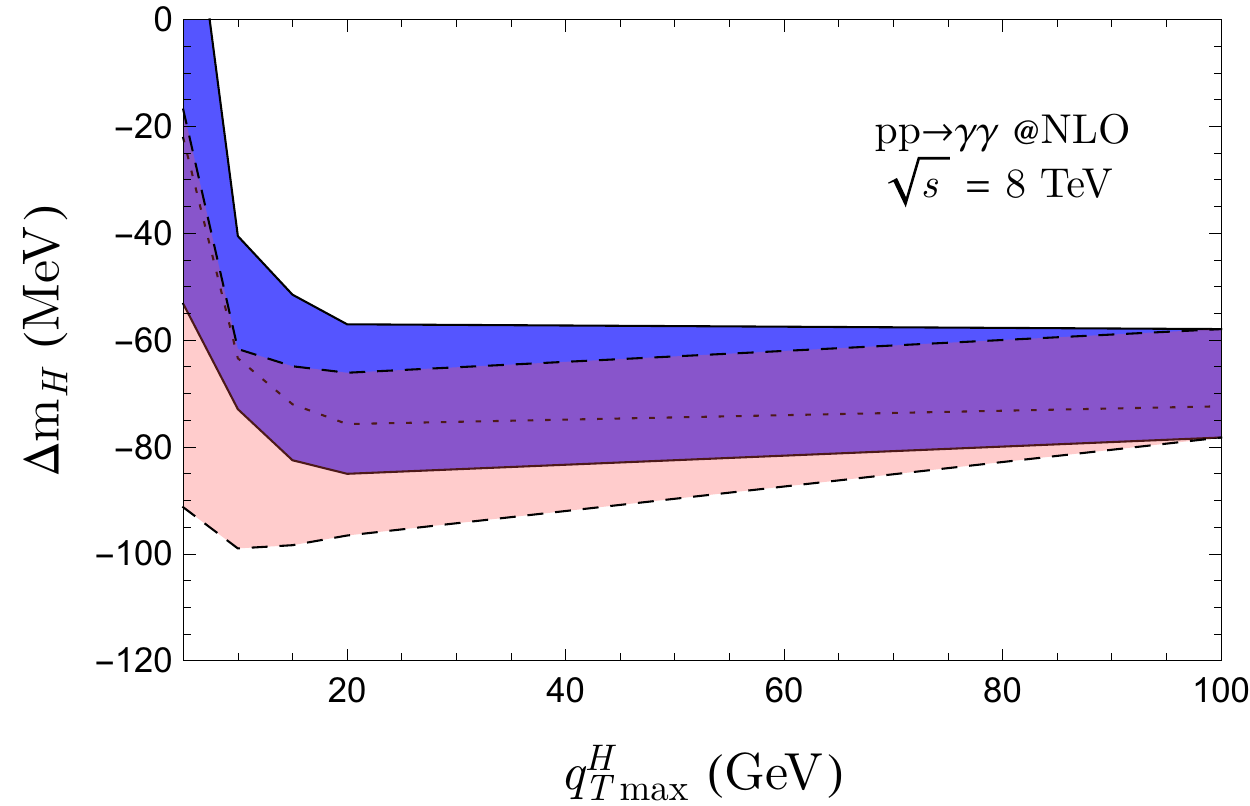}
\caption{\label{fig:jetVeto} \small{Shift in the position of the invariant mass peak $\Delta m_H$ in terms of the maximum Higgs transverse-momentum $q_{T  max}^H$ for the calculation at NLO+NLL. The dark blue band shows our resummed prediction at NLO+NLL, with the dotted line corresponding to the central scale choice $\mu_R=\mu_F=2\mures=m_H$, and the band arising from variation of the scales as described at the beginning of this Section. The light red band gives the corresponding prediction at fixed order, without resummation.}}
\end {center}
\end{figure}

We now turn to the mass shift $\Delta m_H$ induced by the real part of the interference $R$, which is analyzed in figures~\ref{fig:jetVeto} and~\ref{fig:pTmin}. Again, our main goal is to emphasize the relative importance of the contributions from the high and low $q_T$ regions as well as evaluating the impact of resummation on the results.
In figure~\ref{fig:jetVeto} we present the shift $\Delta m_H$ as a function of the kinematic cut $q_{T  max}^H$ already introduced for the plot in figure~\ref{fig:destructive}.
We compare the prediction including resummation (dark blue band) to the fixed order one (light red band). For relatively high transverse momentum ($q_{T  max}^H > 20-30$ GeV) values of $q_{T  max}^H$, the two predictions are consistent (and both are consistent with previously published ones, in particular in ref.~\cite{Dixon:2013haa}) and show a very weak dependence of $\Delta m_H$ on the cut. Again, the two prediction begin to differ significantly (though not so much as in the total cross-section case) in the lower $q_{T  max}^H$ region\footnote{Fig.~\ref{fig:jetVeto} is our direct counterpart to fig.~3 in ref.~\cite{Dixon:2013haa}. Notice that the kinematical cuts are similar but not equivalent in the two analyses, as stated at the beginning of this Section.}.

It is interesting to note that the shift in fact becomes significantly smaller for $q_{T  max}^H \lesssim 5-10$ GeV, eventually becoming compatible with zero inside the range of uncertainty. This could, in principle, allow for an experimental detection of the shift, if a precise enough value of $m_H$ can be extracted experimentally by only using $\gg$ events with small $q_T$. The decrease of $\Delta m_H$ is a manifestation of the fact that the ratio between the real interference contribution $R$ and the signal one $S$ becomes smaller in the low $q_T$ region.
\begin{figure}[!hbt] 
\begin {center}
\hspace*{-0.65cm}
\includegraphics[width=0.45\textwidth]{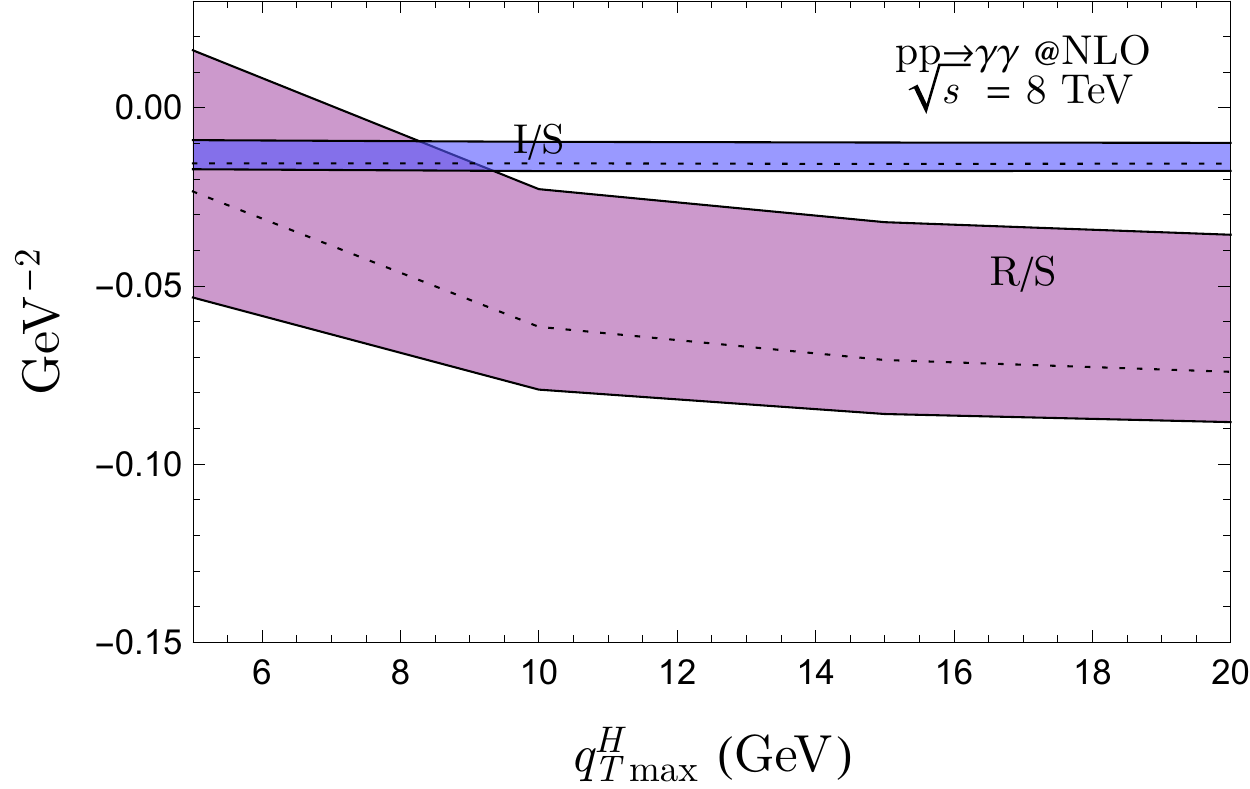}
\caption{\label{fig:RISrate} \small{$R/S$ (purple band) and $I/S$ rates in terms of the maximum Higgs transverse-momentum $q_{T  max}^H$ for the calculation at NLO+NLL. The dotted line represents the result corresponding to the central scale choice $\mu_R=\mu_F=2\mures=m_H$, and the bands show the theoretical uncertainties which arise from variation of the scales as described at the beginning of this Section.}}
\end {center}
\end{figure}
We show explicitly that this is the case in figure~\ref{fig:RISrate}, where we plot the dependence of the ratios $R/S$ and $I/S$ on $q_{T  max}^H$. Notice in particular the contrasting behaviour of the two interference contributions: while $R/S$ drops, the $I/S$ rate remains essentially stable, which leads to the $q_{T max}^H$-independence of the interference suppression to the total cross-section already shown in figure~\ref{fig:destructive}.

\begin{figure}[!hb] 
\begin {center}
\hspace*{-0.65cm}
\includegraphics[width=0.48\textwidth]{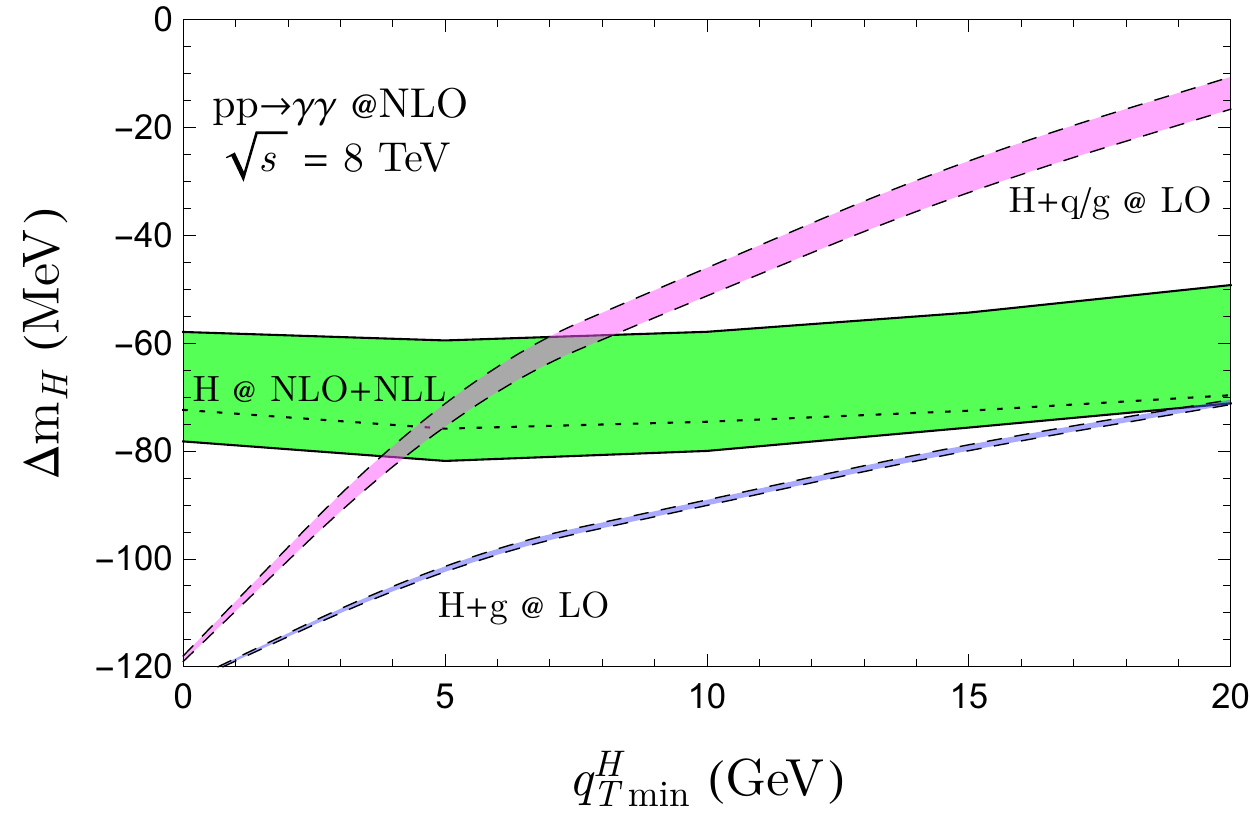}
\caption{\label{fig:pTmin} \small{Shift in the position of the invariant mass peak $\Delta m_H$ in terms of the minimum Higgs transverse-momentum $q_{T  min}^H$ for the calculation at NLO+NLL. The large green band gives our resummed prediction at NLO+NLL, with dotted line corresponding to the central scale choice $\mu_R=\mu_F=2\mures=m_H$, and the the band arising from variation of the scales as described at the beginning of this Section. The thin magenta and blue bands give the corresponding fixed order predictions, the first one including all channels ($gg$, $qg$, $q\bar{q}$) and the latter one -- which we show here for reference -- just the $gg$ channel}}
\end {center}
\end{figure}

Finally, in figure~\ref{fig:pTmin} we plot the resummed prediction (large green band) for $\Delta m_H$ arising from the opposite $q_T$ region, that is by imposing an additional cut on \emph{minimum} transverse-momentum $q_T$ of the diphoton pair (of the Higgs). In this case, the difference in comparison with the fixed order calculation (purple and blue bands in figure~\ref{fig:pTmin}) is dramatic. The substantial improvement brought about by resummation in this case can be understood by the fact that in the fixed order calculation -- since virtual and Born contributions are both namely at $q_T=0$ GeV -- is \emph{effectively leading order} in the region $q_T>0$, that is, the calculation only takes into account contributions with real QCD radiation. 
By contrast, in the resummed result, the Born and virtual corrections are automatically (and physically) spread also to the region $q_T>0$. Much as in the case of the $q_T$ spectrum (figure~\ref{fig:pTspectrumInterf}), the fixed order prediction not only results in a different central value, but also in an artificially small uncertainty band. Of course, this is only true if the $q_{T min}$ cut is moderate: as $q_{T min}$ is increased, eventually there is no more contribution from resummation, and the fixed order prediction must be recovered. By comparison with figure~\ref{fig:pTspectrumInterf}, we can expect this to happen for $q_{T min}> 80$ GeV, way beyond the region shown in the plot.

We can see that in this case the cut has a very low impact on the mass shift, which again remains stable around its average NLO value of $\sim 70$ MeV. (thin magenta and blue bands)

 Finally, it is worth noting the consistence of the resummed predictions from the rightmost part of fig.~\ref{fig:jetVeto} and the leftmost one of fig.~\ref{fig:pTmin}, both of which tend to the total NLO $\Delta m_H$ prediction in the fully inclusive case with no cut on $q_T$, testifying for the robustness of the resummed calculation.

The results of figures~\ref{fig:jetVeto} and~\ref{fig:pTmin} considered together show that the mass shift is a rather \emph{robust} effect, which reinforces the hope that it could be eventually detected experimentally. At the very least the mass shift -- particularly if a more precise value can be obtained using a realistic modelling of experimental effects -- can be used to constrain deviations of the Higgs width from the SM predicted value~\cite{Dixon:2013haa} in an essentially model-independent way. This possibility looks particularly promising if the mass shift constraint coming from the $R$ part of the interference is analyzed together to the complementary one arising from the $I$ term as proposed recently in ref.~\cite{Campbell:2017rke}.

\section{Conclusions}

We presented an upgraded calculation for the effects of interference between the resonant process $pp \to H \to \gamma\gamma$ and the continuum background at LHC. Our calculation includes for the first time all LO and dominant NLO effects (that is, up to $\mathcal{O}(\as^3)$) as well as resummation of large logarithmic $\sim \as^n \ln(q_T^2/Q^2)^n$ terms up to NLL. The interference contribution is as usual split in two parts, one proportional to the imaginary part of the Higgs Breit-Wigner propagator (symmetric around $m_H$), and the other proportional to the real part (antisymmetric). The two terms have complementary effects on physical distributions, the first one modifying the total cross-section and the $q_T$ spectrum, the second one causing a small shift of the measured Higgs mass away from the theoretical input value. We studied the importance of both effects in the high and low Higgs transverse-momentum regions, the latter requiring resummation in order to get reliable results. We obtain a small suppression (of order $1-1.5\%$) of the cross-section due to the imaginary interference contribution which is spread almost evenly along the transverse-momentum spectrum, with the low transverse-momentum (up around 20 GeV) region being slightly more important. By contrast, the mass shift due to real interference is mostly driven by the moderate and high $q_T$ region $q_T > 15$ GeV). In this region, the shift value appears stable at a value of around $70$ MeV, at around $30\%$ less than the LO estimate consistently with what was found in previously published results~\cite{Dixon:2013haa}. While small, this deviation does not appear negligible, being of the same order as the full uncertainty on the best $m_H$ experimental estimate currently available~\cite{Aad:2015zhl}, while the suppression of the shift at low $q_T$ could potentially be used to detect the effect without having to do comparison with different channels, such as ZZ (which would be useful to constrain any deviation of the Higgs width from the SM prediction~\cite{Dixon:2013haa}), if enough statistics at low $q_T$ could be accumulated. Finally, we note that a detailed analysis of the combined effects of the real and imaginary parts of the interference could in principle lead to powerful model-independent constraints on deviations of $\Gamma_H$ from the predicted SM value.

\section*{Acknowledgements}
We would like to thank Stefano Catani for valuable discussions and comments.


\begin{thebibliography}{10}

\bibitem{Chatrchyan:2012xdj}
{\bf CMS} Collaboration, S.~Chatrchyan et~al., {\em Phys. Lett.} {\bf B716}
  (2012) 30--61, [\href{http://arxiv.org/abs/1207.7235}{{\tt
  arXiv:1207.7235}}].

\bibitem{Aad:2012tfa}
{\bf ATLAS} Collaboration, G.~Aad et~al., {\em Phys. Lett.} {\bf B716} (2012)
  1--29, [\href{http://arxiv.org/abs/1207.7214}{{\tt arXiv:1207.7214}}].

\bibitem{Dicus:1987fk}
D.~A. Dicus and S.~S.~D. Willenbrock, {\em Phys. Rev.} {\bf D37} (1988) 1801.

\bibitem{Dixon:2003yb}
L.~J. Dixon and M.~S. Siu, {\em Phys. Rev. Lett.} {\bf 90} (2003) 252001,
  [\href{http://arxiv.org/abs/hep-ph/0302233}{{\tt hep-ph/0302233}}].

\bibitem{Campbell:2017rke}
J.~Campbell, M.~Carena, R.~Harnik, and Z.~Liu,
  \href{http://arxiv.org/abs/1704.08259}{{\tt arXiv:1704.08259}}.

\bibitem{Martin:2012xc}
S.~P. Martin, {\em Phys.Rev.} {\bf D86} (2012) 073016,
  [\href{http://arxiv.org/abs/1208.1533}{{\tt arXiv:1208.1533}}].

\bibitem{deFlorian:2013psa}
D.~de~Florian, N.~Fidanza, R.~Hern\'andez-Pinto, J.~Mazzitelli,
  Y.~Rotstein~Habarnau, and G.~Sborlini, {\em Eur.Phys.J.} {\bf C73} (2013)
  2387, [\href{http://arxiv.org/abs/1303.1397}{{\tt arXiv:1303.1397}}].

\bibitem{Martin:2013ula}
S.~P. Martin, {\em Phys.Rev.} {\bf D88} (2013), no.~1 013004,
  [\href{http://arxiv.org/abs/1303.3342}{{\tt arXiv:1303.3342}}].

\bibitem{Aad:2015zhl}
{\bf ATLAS, CMS} Collaboration, G.~Aad et~al., {\em Phys. Rev. Lett.} {\bf 114}
  (2015) 191803, [\href{http://arxiv.org/abs/1503.07589}{{\tt
  arXiv:1503.07589}}].

\bibitem{Dixon:2013haa}
L.~J. Dixon and Y.~Li, {\em Phys.Rev.Lett.} {\bf 111} (2013) 111802,
  [\href{http://arxiv.org/abs/1305.3854}{{\tt arXiv:1305.3854}}].

\bibitem{Becot:2015xzw}
C.~P. Becot, {\em {Diphoton lineshape of the BEH boson using the ATLAS detector
  at the LHC}}.
\newblock PhD thesis, Orsay, LAL, 2015-09-01.

\bibitem{Catani:2007vq}
S.~Catani and M.~Grazzini, {\em Phys. Rev. Lett.} {\bf 98} (2007) 222002,
  [\href{http://arxiv.org/abs/hep-ph/0703012}{{\tt hep-ph/0703012}}].

\bibitem{Anastasiou:2004xq}
C.~Anastasiou, K.~Melnikov, and F.~Petriello, {\em Phys. Rev. Lett.} {\bf 93}
  (2004) 262002, [\href{http://arxiv.org/abs/hep-ph/0409088}{{\tt
  hep-ph/0409088}}].

\bibitem{Anastasiou:2005qj}
C.~Anastasiou, K.~Melnikov, and F.~Petriello, {\em Nucl. Phys.} {\bf B724}
  (2005) 197--246, [\href{http://arxiv.org/abs/hep-ph/0501130}{{\tt
  hep-ph/0501130}}].

\bibitem{deFlorian:2011xf}
D.~de~Florian, G.~Ferrera, M.~Grazzini, and D.~Tommasini, {\em JHEP} {\bf 11}
  (2011) 064, [\href{http://arxiv.org/abs/1109.2109}{{\tt arXiv:1109.2109}}].

\bibitem{Catani:2011qz}
S.~Catani, L.~Cieri, D.~de~Florian, G.~Ferrera, and M.~Grazzini, {\em Phys.
  Rev. Lett.} {\bf 108} (2012) 072001,
  [\href{http://arxiv.org/abs/1110.2375}{{\tt arXiv:1110.2375}}]. [Erratum:
  Phys. Rev. Lett.117,no.8,089901(2016)].

\bibitem{Campbell:2016yrh}
J.~M. Campbell, R.~K. Ellis, Y.~Li, and C.~Williams, {\em JHEP} {\bf 07} (2016)
  148, [\href{http://arxiv.org/abs/1603.02663}{{\tt arXiv:1603.02663}}].

\bibitem{Cieri:2015rqa}
L.~Cieri, F.~Coradeschi, and D.~de~Florian, {\em JHEP} {\bf 06} (2015) 185,
  [\href{http://arxiv.org/abs/1505.03162}{{\tt arXiv:1505.03162}}].

\bibitem{Djouadi:1997yw}
A.~Djouadi, J.~Kalinowski, and M.~Spira, {\em Comput. Phys. Commun.} {\bf 108}
  (1998) 56--74, [\href{http://arxiv.org/abs/hep-ph/9704448}{{\tt
  hep-ph/9704448}}].

\bibitem{Bern:1993mq}
Z.~Bern, L.~J. Dixon, and D.~A. Kosower, {\em Phys. Rev. Lett.} {\bf 70} (1993)
  2677--2680, [\href{http://arxiv.org/abs/hep-ph/9302280}{{\tt
  hep-ph/9302280}}].

\bibitem{Bern:1994fz}
Z.~Bern, L.~J. Dixon, and D.~A. Kosower, {\em Nucl. Phys.} {\bf B437} (1995)
  259--304, [\href{http://arxiv.org/abs/hep-ph/9409393}{{\tt hep-ph/9409393}}].

\bibitem{Spira:1995rr}
M.~Spira, A.~Djouadi, D.~Graudenz, and P.~M. Zerwas, {\em Nucl. Phys.} {\bf
  B453} (1995) 17--82, [\href{http://arxiv.org/abs/hep-ph/9504378}{{\tt
  hep-ph/9504378}}].

\bibitem{Bern:2001df}
Z.~Bern, A.~De~Freitas, and L.~J. Dixon, {\em JHEP} {\bf 09} (2001) 037,
  [\href{http://arxiv.org/abs/hep-ph/0109078}{{\tt hep-ph/0109078}}].

\bibitem{Shifman:1979eb}
M.~A. Shifman, A.~I. Vainshtein, M.~B. Voloshin, and V.~I. Zakharov, {\em Sov.
  J. Nucl. Phys.} {\bf 30} (1979) 711--716. [Yad. Fiz.30,1368(1979)].

\bibitem{Dawson:1990zj}
S.~Dawson, {\em Nucl. Phys.} {\bf B359} (1991) 283--300.

\bibitem{Djouadi:1991tka}
A.~Djouadi, M.~Spira, and P.~M. Zerwas, {\em Phys. Lett.} {\bf B264} (1991)
  440--446.

\bibitem{Dawson:1993qf}
S.~Dawson and R.~Kauffman, {\em Phys. Rev.} {\bf D49} (1994) 2298--2309,
  [\href{http://arxiv.org/abs/hep-ph/9310281}{{\tt hep-ph/9310281}}].

\bibitem{Bern:2002jx}
Z.~Bern, L.~J. Dixon, and C.~Schmidt, {\em Phys. Rev.} {\bf D66} (2002) 074018,
  [\href{http://arxiv.org/abs/hep-ph/0206194}{{\tt hep-ph/0206194}}].

\bibitem{Bozzi:2005wk}
G.~Bozzi, S.~Catani, D.~de~Florian, and M.~Grazzini, {\em Nucl. Phys.} {\bf
  B737} (2006) 73--120, [\href{http://arxiv.org/abs/hep-ph/0508068}{{\tt
  hep-ph/0508068}}].

\bibitem{Catani:2013tia}
S.~Catani, L.~Cieri, D.~de~Florian, G.~Ferrera, and M.~Grazzini, {\em Nucl.
  Phys.} {\bf B881} (2014) 414--443,
  [\href{http://arxiv.org/abs/1311.1654}{{\tt arXiv:1311.1654}}].

\bibitem{Bozzi:2007pn}
G.~Bozzi, S.~Catani, D.~de~Florian, and M.~Grazzini, {\em Nucl. Phys.} {\bf
  B791} (2008) 1--19, [\href{http://arxiv.org/abs/0705.3887}{{\tt
  arXiv:0705.3887}}].

\bibitem{Kodaira:1981nh}
J.~Kodaira and L.~Trentadue, {\em Phys. Lett.} {\bf B112} (1982) 66.

\bibitem{Catani:1988vd}
S.~Catani, E.~D'Emilio, and L.~Trentadue, {\em Phys. Lett.} {\bf B211} (1988)
  335--342.

\bibitem{Davies:1984hs}
C.~T.~H. Davies and W.~J. Stirling, {\em Nucl. Phys.} {\bf B244} (1984)
  337--348.

\bibitem{Davies:1984sp}
C.~T.~H. Davies, B.~R. Webber, and W.~J. Stirling, {\em Nucl. Phys.} {\bf B256}
  (1985) 413. [1,I.95(1984)].

\bibitem{deFlorian:2000pr}
D.~de~Florian and M.~Grazzini, {\em Phys. Rev. Lett.} {\bf 85} (2000)
  4678--4681, [\href{http://arxiv.org/abs/hep-ph/0008152}{{\tt
  hep-ph/0008152}}].

\bibitem{Catani:1989ne}
S.~Catani and L.~Trentadue, {\em Nucl. Phys.} {\bf B327} (1989) 323--352.

\bibitem{Becher:2010tm}
T.~Becher and M.~Neubert, {\em Eur. Phys. J.} {\bf C71} (2011) 1665,
  [\href{http://arxiv.org/abs/1007.4005}{{\tt arXiv:1007.4005}}].

\bibitem{Frixione:1998jh}
S.~Frixione, {\em Phys. Lett.} {\bf B429} (1998) 369--374,
  [\href{http://arxiv.org/abs/hep-ph/9801442}{{\tt hep-ph/9801442}}].

\bibitem{Martin:2009iq}
A.~Martin, W.~Stirling, R.~Thorne, and G.~Watt, {\em Eur.Phys.J.} {\bf C63}
  (2009) 189--285, [\href{http://arxiv.org/abs/0901.0002}{{\tt
  arXiv:0901.0002}}].

\end{thebibliography}

\providecommand{\href}[2]{#2}\begingroup\raggedright\endgroup

\end{document}